\documentclass[a4paper]{article}
\usepackage{INTERSPEECH_v2}
\usepackage{algorithm2e}
\usepackage{float}
\usepackage{tabularx}
\usepackage{booktabs}
\usepackage{psfrag}    
\usepackage{amssymb,amsmath,epsfig}
\usepackage{subfigure}
\floatstyle{ruled}
\newfloat{algorithm}{htbp}{loa}
\floatname{algorithm}{Algorithm}

\usepackage{xcolor}
\title{\vspace{-1cm}Tied Hidden Factors in Neural Networks for End-to-End Speaker Recognition}

\name{Antonio Miguel, Jorge Llombart, Alfonso Ortega, Eduardo Lleida}
\address{ViVoLAB, Aragon Institute for Engineering Research (I3A), University of Zaragoza, Spain}
\email{\{amiguel,jllombg,ortega,lleida\}@unizar.es}

\begin{document}

\maketitle
\begin{abstract}
In this paper we propose a method to model speaker and session variability and able to generate likelihood ratios using neural networks in an end-to-end phrase dependent speaker verification system. 
As in Joint Factor Analysis, the model uses tied hidden variables to model speaker and session variability and a MAP adaptation of some of the parameters of the model.
In the training procedure our method jointly estimates the network parameters and the values of the speaker and channel hidden variables. 
This is done in a two-step backpropagation algorithm, first the network weights and factor loading matrices are updated and then the hidden variables, whose gradients are calculated by aggregating the corresponding speaker or session frames, since these hidden variables are tied. 
The last layer of the network is defined as a linear regression probabilistic model whose inputs are the previous layer outputs. 
This choice has the advantage that it produces likelihoods and additionally it can be adapted during the enrolment using MAP without the need of a gradient optimization. 
The decisions are made based on the ratio of the output likelihoods of two neural network models, speaker adapted and universal background model.
The method was evaluated on the RSR2015 database.

\end{abstract}
\footnote{\textcolor{red}{Cite as: A. Miguel, J. Llombart, A. Ortega, E. Lleida, ``Tied Hidden Factors in Neural Networks for End-to-End Speaker Recognition.''
 In Proc. Interspeech, pp 2819--2823, Stockholm, Sweden, 2017.}}

\section{Introduction}

Deep neural networks (DNNs) have been successfully applied in many speech and speaker recognition tasks in recent years, providing outstanding performances.
In speech technologies most of the DNN solutions have used them as classifiers or feature extractors, but in this work we propose to apply them in an end-to-end detection task, where the output of the system is a likelihood score ratio, which after applying a threshold provides good performance without the need of further calibration.
Unfortunately, the high number of parameters of DNNs and their tendency to overfit data make that type of detection task difficult to approach.
In this work we try to take advantage of the high flexibility of these models and their capacity to learn nonlinear patterns from the input signals.
This has required to provide solutions to decrease the overfitting problems and their lack of a measure of uncertainty by adding external control variables to model session and speaker variability, and also proposing Bayesian adaptation and evaluation mechanisms.

Recent approaches to speaker recognition use Joint Factor Analysis (JFA) \cite{Yin2007,Kenny2008,kenny2014joint,Miguel2014,Stafylakis2016SpeakerRecognition} with GMMs or HMMs as base distributions, i-vector systems \cite{Dehak2010,villalba-bay2cov-is2011,Stafylakis2013}, neural networks as bottleneck feature extractors for JFA or i-vector systems \cite{Liu2015DeepVerification,zeinali2016deep}, or neural networks as classifiers to produce posterior probabilities for JFA or i-vector extractors \cite{Dey2016DeepVerification,zeinali2016deep,zeinali2016vector}.
There have been other approaches to create speaker verification using neural networks by means of LSTM networks \cite{Moreno2016} to provide sequence to vector compression or to extract total variability latent factors (similar to i-vector) directly from a neural network in \cite{garimella2013factor}, in that case a PLDA backend was used in a text independent speaker recognition task. 
The proposed method has several parallelisms to JFA models since we also encode speaker and session information using latent variables, and the model probability distribution can also be adapted to the speaker data using MAP or Bayesian techniques, but in this paper the model is built on top of an autoencoder neural network as an alternative to other models like GMMs or HMMs.
The proposed method is an end-to-end solution since the neural network performs all the processing steps and it provides the likelihood ratio.
The autoencoder \cite{kingma2013auto} is a generative model that is trained to predict its input with the restriction that the network has a bottleneck layer of smaller dimension. 
Its training is unsupervised  in terms of frame level phoneme labels, what makes it a candidate to substitute GMMs as the underlying model of the system. 
In addition, it can be robustly trained using Bayesian methods and its output layer can be probabilistic, as it has been shown in recent works \cite{kingma2013auto,kingma2015variational,blundell2015weight}. 
As we show later in the paper, we can build a system using these probabilistic autoencoders, but performance can be improved by using tied hidden variables to model speaker and session variability.
Speaker factors have been used as a source of additional information for neural networks in speech recognition to improve the performance of speaker independent recognizers \cite{Moreno2014,xue2014fast}.
In many cases the latent variables were obtained by using an external system like a GMM based JFA or i-vector extractor, and the network is then retrained to capture this extra information \cite{samarakoon2016factorized}.
There have been works were speaker factors or speaker codes are used to enhance speech recognition systems and they were optimized by gradient optimization \cite{samarakoon2016factorized,huang2016speaker} or initialized using Singular Value Decomposition \cite{xue2015unsupervised}. 
We propose a joint factor approach for neural networks to model speaker and session variability, showing that effective improvements can be obtained by using the latent factors with respect to a reference network.
%
%
A modified two-step backpropagation algorithm is proposed to train the model parameters and the latent factors, first the network weights and factor loading matrices are updated given the current value of the latent variables and then the latent variables are updated.
To calculate the gradients of the cost function with respect to the network weights the minibatch samples can be randomly permuted to facilitate convergence, but the gradients with respect to the hidden factors are calculated by aggregating all the corresponding speaker or session frames, since these hidden variables are tied. 
%

%
This paper is organized as follows.
In Section \ref{s2} the tied factor analysis model for neural networks is presented.
Section \ref{s3} discusses the use of autoencoder neural networks for speaker verification. 
Section \ref{s4} presents an experimental study to show the accuracy of the models in phrase dependent speaker recognition.
Conclusions are presented in Section \ref{s5}.

\section{Tied Hidden Factors in Neural Networks}
\label{s2}

\begin{figure}[tb]
\begin{center}
   	\includegraphics[width=0.9\columnwidth]{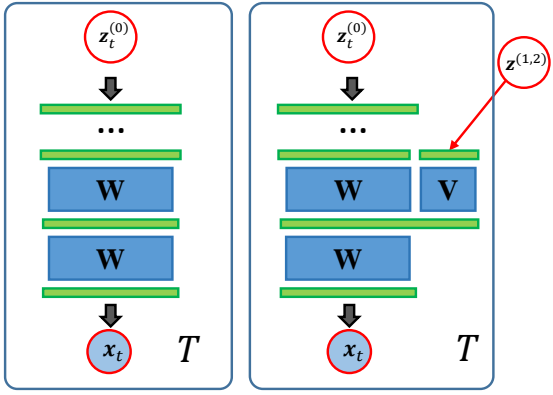} 
\end{center}
\caption{ Decoder in an autoencoder for DNNs and TF-DNN}
\label{fig1}
\end{figure} 

The concept of tied hidden factors to model contextual or sequence information has appeared in many different contexts like face recognition \cite{Prince2008TiedDifferences}, speech recognition \cite{Kuhn2000,Kenny2005a}, language recognition \cite{Martinez2011a,Martinez2013a}, speaker diarization \cite{Vaquero2013} or audio segmentation \cite{Castan2013b}, but their most prominent field has been speaker recognition with JFA or i-vector approaches  \cite{Kenny2006,Kenny2008,kenny2014joint,Miguel2014,Stafylakis2016SpeakerRecognition,Dehak2010} 
The use of these type of global variables has been defined in more general approaches as hierarchical models \cite{gelman2006data} and more recently in the context of DNNs in \cite{Tran2017}. 
We propose to use two types of tied hidden factors in neural networks to extend previous works with a general algorithm to estimate them.
We refer to this model in general as Tied Factor Deep Neural Network (TF-DNN), and for the special case of two factors speaker and session, TF2-DNN.
%
%
In Figure \ref{fig1} it is depicted the conceptual difference of the decoder part of a standard DNN autoencoder and a TF-DNN, where $\mathbf{z}_t^{(0)}$ is the bottleneck layer which changes for every frame $t$, but there are tied variables that affect the output of many samples $\mathbf{x}_t$ of the same session (or file in databases like RSR2015) $\mathbf{z}_f^{(1)}$ and same speaker $\mathbf{z}_s^{(2)}$, displaying a similar the idea to \cite{kenny2014joint,Miguel2014} for GMMs.

To define the model, first we need to describe the observed data, $\mathbf{X} = \{\mathbf{x}_t\}_{t=1}^T$ as a sequence of feature vectors $\mathbf{x}_t \in \mathbb{R}^D$ with $D$ the feature dimension. 
In the proposed TF2-DNN approach we assume that a set of hidden variables encode speaker and session information.
The session and speaker latent factors are denoted as $\mathbf{Z}^{(1)} = \{\mathbf{z}^{(1)}_f\}_{f=1}^F$ and $\mathbf{Z}^{(2)} = \{\mathbf{z}^{(2)}_s\}_{s=1}^S$ with $\mathbf{z}^{(1)}_f \in \mathbb{R}^{R^{(1)}}$ and $\mathbf{z}^{(2)}_s \in  \mathbb{R}^{R^{(2)}}$, where $F$ and $S$ are the number of sessions and speakers and $R^{(1)}$ and $R^{(2)}$ are the dimension of their respective subspaces. 
The complete set of hidden variables is denoted as  $\mathcal{Z} = (\mathbf{Z}^{(1)}, \mathbf{Z}^{(2)})$, they encode speaker and session information, and since they are unknown we need to estimate them using a labeled dataset.
The training data are typically organized by sessions. 
If sessions are labeled by speaker, then it is straightforward to obtain frame level session labels $\phi^{(1)}_t \in \{ 1,\ldots,F\}$, and speaker labels as $\phi^{(2)}_t \in \{ 1,\ldots,S\}$, so that the training dataset is defined as $\mathcal{D}=\{\mathbf{x}_t, \phi^{(1)}_t, \phi^{(2)}_t \}_{t=1}^T$, for each data sample we need its speaker and session label.

The TF2-DNN is built on top of a regular neural network whose parameters are the weights $\mathbf{W}_l$ and biases $\mathbf{b}_l$ of all the layers $l=1,\ldots,L$. 
We denote them by $\bar{\mathbf{W}} = \{\mathbf{W}_1, \mathbf{b}_1, \ldots, \mathbf{W}_L, \mathbf{b}_L\}$ for a NN with $L$ layers.
The layers of the network which are connected to the latent variables are called TF2 layers and have additional parameters and a different output than a standard network.
In Figure \ref{fig1}, we show a simplified model of a DNN with standard layers and TF-DNN with a TF2 layer. 
In the case of linear embedding, a factor loading matrix $\mathbf{V}_l$ is required for each factor and the layer $l$ output is defined as
\begin{eqnarray}
\label{eq:1}
	\mathbf{x}_{t,l} =
    \sigma( \mathbf{W}_l \mathbf{x}_{t,l-1} + \mathbf{b}_l + \mathbf{V}^{(1)}_l \mathbf{z}^{(1)}_f + \mathbf{V}^{(2)}_l \mathbf{z}^{(2)}_s ),
\end{eqnarray}
where $\mathbf{x}_{t,l-1}$ and $\mathbf{x}_{t,l}$ are the previous layer output and the current layer output, the function $\sigma()$ is the layer nonlinearity, $\mathbf{z}^{(1)}_f$ is the session factor corresponding to the frame $\mathbf{x}_{t}$, and $f$ is the corresponding file label $f=\phi^{(1)}_t$, and $\mathbf{z}^{(2)}_s$ is the corresponding speaker factor and $s$ is the speaker label $s=\phi^{(2)}_t$.
The set of all the network parameters is denoted as $\mathbf{\Theta} = (\bar{\mathbf{W}}, \bar{\mathbf{V}})$.


\begin{algorithm}[t]
\SetAlgoLined
\KwIn{$\mathcal{D}$: Acoustic features $\mathbf{X}$, frame level session labels and frame level speaker labels $\phi^{(1)},\phi^{(2)}$, and prior values for the initialization $\mathbf{\lambda}$, learning rate values for the unknown parameters $\mathbf{\alpha}$, and number of epochs $N$ }
\KwOut{Estimations for hidden factors $\mathcal{Z} = (\mathbf{Z}^{(1)}, \mathbf{Z}^{(2)})$ and the network parameters $\mathbf{\Theta} = (\bar{\mathbf{W}}, \bar{\mathbf{V}})$: the neural network weighs and factor loading matrices }

\vspace{0.1cm}

\textbf{1. Initialization}\;

Initialize all the unknown weights and latent factors randomly following their prior distribution:\;

$\mathbf{\Theta} \sim \mathcal{N}(\mathbf{0}, \lambda_1 \mathbf{I})$,\;
$\mathbf{z}^{(1)} \sim \mathcal{N}(\mathbf{0}, \lambda_2 \mathbf{I}),$\;
$\mathbf{z}^{(2)} \sim \mathcal{N}(\mathbf{0}, \lambda_3 \mathbf{I})$\;

\vspace{0.2cm}
\textbf{2. Two step backpropagation}\;

\For{ $n \leftarrow 1$ \KwTo $N$ }{
	    
    \textbf{2.1 Step 1, backpropagation parameters: $\mathbf{\Theta}$:}\
    
    Minibatch $b$ updates: frames $t_b$ are selected randomly:\;
       
    $\mathcal{D}_{b} \leftarrow \{\mathbf{x}_t, \phi^{(1)}_t, \phi^{(2)}_t | t \in t_b \}$,
   
   $ \mathcal{Z}_{b} 
    \leftarrow \{   
    \mathbf{z}^{(1)}_{f}, 
    \mathbf{z}^{(2)}_{s} | t \in t_b, f = \phi^{(1)}_t, s = \phi^{(2)}_t\}$
        
    $\mathbf{\Theta} \leftarrow 
    \mathbf{\Theta} - \alpha_1 \nabla_{\mathbf{\Theta}} J(\mathcal{D}_{b},\mathcal{Z}_{b}, \mathbf{\Theta})$\;
    
     \textbf{2.2 Step 2, backpropagation hidden variables $\mathbf{z}^{(1)}$, $\mathbf{z}^{(2)}$ }\;
     
     Using expressions \eqref{eq:2}, \eqref{eq:3} for all speaker $s$ and sessions $f$

    $\mathbf{z}^{(1)}_f \leftarrow 
    \mathbf{z}^{(1)}_f - \alpha_2 \nabla_{\mathbf{z}^{(1)}_f} J(\mathcal{D},\mathcal{Z}, \mathbf{\Theta} )$\;

    $\mathbf{z}^{(2)}_s \leftarrow 
    \mathbf{z}^{(2)}_s - \alpha_3 \nabla_{\mathbf{z}^{(2)}_s} J(\mathcal{D},\mathcal{Z}, \mathbf{\Theta} )$\;
    
}

\caption{Training algorithm for two tied hidden factor neural network (TF2-DNN)}
\label{alg:1}
\end{algorithm}


Given a cost function $J(\mathcal{D},\mathcal{Z},\mathbf{\Theta})$ that can be evaluated for some training data, $\mathcal{D}$, and an instance of the unknown parameters $\mathcal{Z}$ and $\mathbf{\Theta}$, we can define an optimization method to minimize the cost.
In this work we have used gradient based optimization since it can be scaled to larger datasets and still be tractable.
To estimate both the network weights and the tied latent factors Algorithm \ref{alg:1} is proposed.
This algorithm optimizes both sets of variables in an alternate way like in coordinate descent type algorithms.
In \cite{Miguel2014} the alternate E and M steps had the same motivation, since the E step considered the likelihood as the objective and the latent variables were optimized in a search process given the other parameters fixed.
The gradients with respect to the network parameters $\mathbf{\Theta}$ in the step $2.1$ are computed as usual gradients since the hidden variables are given as argument with their current value and they can be interpreted as external information to the network.
The gradients with respect to the tied hidden variables have to be considered more carefully, since they have to be calculated by aggregating all the corresponding speaker or session frames since they are tied.
Then the gradients for session $f$ and speaker $s$ factors are
\begin{eqnarray}
\label{eq:2}
	\nabla_{\mathbf{z}^{(1)}_f} 
    J(\mathcal{D},\mathcal{Z},\mathbf{\Theta} ) = \!\!\!\!
    \sum_{t | \phi^{(1)}_t = f} 
    \nabla_{\mathbf{z}^{(1)}}  
    J(\mathbf{x}_t,\mathbf{z}^{(1)}_{\phi^{(1)}_t},\mathbf{z}^{(2)}_{\phi^{(2)}_t},\mathbf{\Theta} ) \\
\label{eq:3}
    \nabla_{\mathbf{z}^{(2)}_s} 
    J(\mathcal{D},\mathcal{Z}, \mathbf{\Theta}) = \!\!\!\!
    \sum_{t | \phi^{(2)}_t = s} 
    \nabla_{\mathbf{z}^{(2)}}  
    J(\mathbf{x}_t,\mathbf{z}^{(1)}_{\phi^{(1)}_t},\mathbf{z}^{(2)}_{\phi^{(2)}_t},\mathbf{\Theta} ).
\end{eqnarray}

\section{Neural network end-to-end speaker verification system}
\label{s3}
The autoencoder \cite{kingma2013auto,kingma2015variational} is a generative model that is trained to predict its input with the restriction that the network has a bottleneck layer of smaller dimension. 
Then, the system has two parts: the first part, encoder, learns how to compress the information and the second part, decoder, learns how to reconstruct the signal. 
To adapt this type of network to the task of speaker recognition we need to compute likelihoods of the observed data $\mathbf{x}_t$.
In \cite{kingma2013auto} the bottleneck layer of the autoencoder was considered analogous to the hidden variable $\mathbf{z}_t$ in a factor analysis model \cite{GhahramaniVariationalAnalysersb}. 
Then, the encoder part was associated to a variational approximation to the posterior distribution $q(\mathbf{z}_t | \mathbf{x}_t)$, and the decoder part could be associated to the likelihood of the observed data given the hidden variable by parametrizing a probabilistic model using the network outputs $p(\mathbf{x}_t | \mathbf{z}_t)$.
%
%
To follow with the previous section notation for the latent factors, \cite{Miguel2014}, the bottleneck layer of the autoencoder is denoted as $\mathbf{z}^{(0)}_t$, since it encodes intra-frame information \cite{Ghahramani96,GhahramaniVariationalAnalysersb} and it is the lowest level in the hierarchy: frame (0), session (1), speaker (2).
We can see that other levels could be added easily.

\subsection{Linear Regression probability model}
In this paper we use the same parametrization mechanism as in \cite{kingma2013auto,blundell2015weight} to define that the last layer provides the mean vector of a Gaussian distribution, which we combine with the following idea: if the last layer is a linear function,  $\mathbf{x}_{t,L} = \mathbf{W}_L \mathbf{x}_{t,L-1} + \mathbf{b}_L$, the likelihood can also be interpreted as a linear regression probability model whose regression coefficient matrix $\mathbf{B}$ is the last layer weight matrix $\mathbf{W}_L$ as
\begin{equation}
\label{eq:lin}
	p( \mathbf{x}_t |\mathbf{z}_t^{(0)}) = 
    \mathcal{N}( \mathbf{x}_{t,L} , \mathbf{\Psi}  ) 
     = 
    \mathcal{N}( \mathbf{W}_L \mathbf{x}_{t,L-1} + \mathbf{b}_L, \mathbf{\Psi} ),
\end{equation}
where the output $\mathbf{x}_{t,L}$ acts as mean and we can define an arbitrary covariance matrix $\mathbf{\Psi}$.

The following steps could be carried out for the bias parameter $\mathbf{b}_L$ and the covariance matrix $\mathbf{\Psi}$ as well, but to keep the notation simpler \cite{Bishop2008ALearning}, and focus on the most important parameters, the weigths $\mathbf{B} = \mathbf{W}_L$, we derive the network adaptation mechanism for this simpler distribution
\begin{eqnarray}
\label{eq:4}
   p( \mathbf{x}_t | \mathbf{y}_{t})  = \mathcal{N}( \mathbf{B} \mathbf{y}_{t}, \beta^{-1} \mathbf{I} ),
\end{eqnarray}
where $\beta^{-1}\mathbf{I}$ is the covariance matrix, now controlled by a single parameter and we denote the outputs of the previous layer as $\mathbf{y}_t$ for simplicity, with $\mathbf{y}_t =\mathbf{x}_{t,L-1} = f(\mathbf{z}^{(0)}_t)$ using the decoder part of the network except the last layer.

The analogy in expressions \eqref{eq:lin} and \eqref{eq:4} allows to estimate the value of $\mathbf{W}_L$ using a probabilistic approach if we let the rest of the network parameters and hidden variables unchanged, what makes easy to apply ML, MAP or Bayesian estimation techniques.
Given some training data $\mathbf{X}$ organized by rows, and the output previous to the last layer $\mathbf{Y}$, the ML estimator is equivalent to minimize square error, MSE, and is obtained as
\begin{eqnarray}
\label{eq:5}
	\mathbf{B}^{ML} & = & (\mathbf{Y}^\intercal \mathbf{Y})^{-1} \mathbf{Y}^\intercal \mathbf{X}. 
 \end{eqnarray}

The ML estimator can have problems when inverting the matrix if it is ill-conditioned. 
To solve that we can impose a penalty to the weights $\mathbf{B}$ by assuming a Gaussian prior for them $\mathbf{B} \sim \mathcal{N}( \mathbf{B}_0, \lambda_0^{-1} \mathbf{I}  )$, which makes the optimization of the posterior distribution $p(\mathbf{B} | \mathbf{X})$ equivalent to an L2 regularization \cite{blundell2015weight}, this is the MAP estimator
\begin{eqnarray}
\label{eq:6}
	\mathbf{B}^{MAP} & = & (\beta\mathbf{Y}^\intercal \mathbf{Y} +\lambda_0  \mathbf{I})^{-1} ( \beta\mathbf{Y}^\intercal \mathbf{X} + \lambda_0 \mathbf{B}_0),
\end{eqnarray}
which in the case the prior mean is zero it is usually expressed as
\begin{eqnarray}
\label{eq:7}
	\mathbf{B}^{MAP} & = & (\mathbf{Y}^\intercal \mathbf{Y} + \frac{\lambda_0}{\beta}  \mathbf{I})^{-1} \mathbf{Y}^\intercal \mathbf{X}. 
\end{eqnarray}

The fully Bayesian approach \cite{Bishop2008ALearning} provides a posterior distribution for the weights given the priors and the observed data that follows a normal distribution, whose mean has the same value as the MAP estimation in \eqref{eq:6}
\begin{eqnarray}
\label{eq:8}
	p( \mathbf{B} | \mathbf{Y} ) & = & 
    \mathcal{N}( \mathbf{B}_N, \Sigma_N ) \\
 \label{eq:8.1}   
     \mathbf{B}_N & = & \Sigma_N ( \beta \mathbf{Y}^\intercal \mathbf{X} + \lambda_0 \mathbf{B}_0) \\
     \Sigma_N^{-1} & = & (\beta \mathbf{Y}^\intercal \mathbf{Y} +\lambda_0  \mathbf{I}).
\end{eqnarray}

\subsection{UBM training, speaker enrolment and trial evaluation}

Once the building blocks of the model have been established, we describe all the basic steps involved in a speaker recognition system: UBM training, speaker enrolment, and trial evaluation.

The universal background model (UBM) is trained using Algorithm \ref{alg:1} after a random initialization of all the unknown parameters and latent factors.
The labels to assign training frames to the speaker and latent factors must be supplied to the algorithm.
When the UBM is trained we extract the sums
\begin{equation}
\label{eq:9}
 S_{yy}   =  \mathbf{Y}^\intercal \mathbf{Y} = \sum_t \mathbf{y}_t \mathbf{y}_t^\intercal, \,\,\,\,\,
 S_{yx}   =  \mathbf{Y}^\intercal \mathbf{X} = \sum_t \mathbf{y}_t \mathbf{x}_t^\intercal,
\end{equation}
which are the sufficient statistics needed to make adaptations at enrolment time without the need of processing the whole database each time. $\mathbf{y}_t$ and $\mathbf{x}_t$ are column vectors corresponding to frame $t$. Then, $\mathbf{B}^{ubm}$ can be obtained from the stats.

To enrol a speaker in the system in the context of the proposed method requires to create an adapted network to the samples used for enrolment, which are a small number compared to the UBM.
Two mechanism are available in this model for this.
The first one is to adapt the speaker latent factor by using step $2.2$ of the algorithm for a number of iterations using the enrolment data and using as initial values the UBM parameters.
In this case step $2.1$ would not be applied since the network weights have to remain fixed.
The second mechanism is more similar to MAP in JFA systems, two possible adaptations are proposed given the previous linear regression expressions.
One option is to consider the prior mean as the UBM value $\mathbf{B}_0 = \mathbf{B}^{ubm}$, then using expression \eqref{eq:8.1} the maximum of the posterior distribution $p(\mathbf{B}^{spk} | \mathbf{Y}_{spk}, \mathbf{B}_0 = \mathbf{B}^{ubm} )$
\begin{eqnarray}
\label{eq:10}
	\mathbf{B}^{spk} & = & (\beta S_{yy}^{spk} +\lambda_0  \mathbf{I})^{-1} ( \beta S_{yx}^{spk} + \lambda_0 \mathbf{B}^{ubm}).
\end{eqnarray}
Other option is to consider the posterior distribution given both the enrolment and the UBM data $p(\mathbf{B}^{spk} |\mathbf{Y}_{ubm}, \mathbf{Y}_{spk})$, with the prior mean equal to zero. 
To control the weight of the UBM samples with respect to the enrolment we introduce an interpolation factor $\alpha$
\begin{align}
\label{eq:11}
\lefteqn{ 	\mathbf{B}^{spk}  = } \\ 
\nonumber
 & ( \alpha S_{yy}^{ubm} + (1 - \alpha)  S_{yy}^{spk} +  
    								\frac{\lambda_0}{\beta}  \mathbf{I} )^{-1} 
    ( \alpha S_{yx}^{ubm} + (1 - \alpha)  S_{yx}^{spk} ).
\end{align}

Finally for trial evaluation we evaluate the likelihood ratio
\begin{align}
\label{eq:ratio}
\Lambda = \frac{ p( \mathbf{X} | \mathbf{Y}_{ubm}, \mathbf{Y}_{spk} )}
{ p( \mathbf{X} | \mathbf{Y}_{ubm} ) },
\end{align}
where likelihoods are calculated using expression \eqref{eq:lin}, $\mathbf{B}^{ubm}$ in the denominator is estimated using \eqref{eq:8.1} and the sufficient stats, \eqref{eq:9}, for all the UBM data $\mathbf{Y}_{ubm}$.
$\mathbf{B}^{spk}$ in the numerator is estimated using \eqref{eq:11} and the speaker and UBM data, $\mathbf{Y}_{spk}, \mathbf{Y}_{ubm}$, (since it performed better than \eqref{eq:10} in preliminary experiments).

\subsection{Bayesian inference}
Recent advances on applying Bayesian estimation techniques to DNNs  have been shown to be effective against overfitting and to deal with uncertainty \cite{welling2011bayesian,kingma2013auto,blundell2015weight}.
To avoid overfitting when data size is small, we propose to use dropout layers, \cite{srivastava14a}, interleaved with the TFA2 layers, as an alternative to the variational approach \cite{kingma2015variational}.
We train the network using the proposed algorithm with the dropout Bernoulli distribution, $\mathbf{\xi} \sim Be(p)$, which switches off some of the layer outputs with probability $p$.
Then we perform the trial evaluation by sampling
\begin{equation}
 \label{eq:sam}
	 p(\mathbf{x}_t|\mathbf{z}_t^{(0)}) \! =  \!\! \int p(\mathbf{x}_t|\mathbf{z}_t^{(0)}\!\!,\mathbf{\xi}) p(\mathbf{\xi}) d\mathbf{\xi} \\
    \simeq  \frac{1}{L} \!\!\!
    \sum_{\mathbf{\xi}_l \sim Be(p)} p(\mathbf{x}_t|\mathbf{z}_t^{(0)},\mathbf{\xi}_l).
\end{equation}

\vspace{-1mm}

\section{Experiments}
\label{s4}

The experiments have been conducted on the RSR2015 part I text dependent speaker recognition database \cite{Larcher2014}.
The speakers are distributed in three subsets: bkg, dev and eval.
We have only used background data (bkg) to train the UBMs, which can be phrase independent or phrase dependent, as in \cite{Stafylakis2013,kenny2014joint,Miguel2014}.
The evaluation part is used for enrolment and trial evaluation. 
The dev part was not used in these experiments and files were not rejected because of low quality.
Speaker models are build using 3 utterances for each combination of speaker and phrase (1708 for males and 1470 for females).
For testing we have selected the trials using the same phrase as the model, called impostor-correct in \cite{Larcher2014}: 10244(tgt) + 573664(non) = 583908 male trials; 8810(tgt) + 422880(non) = 431690 female trials.
We use the database 16kHz signals to extract 20 MFCCs and their first and second derivatives. 
Then, an energy based voice activity detector is used and data are normalized using short term Gaussianization.
In the experiments in this work the speaker factor is a speaker-phrase combination as in \cite{kenny2014joint}.
To train the autoencoders in this work we used the same DNN architecture in all of the experiments of 4 hidden layers of 500 units, softplus nonlinearities \cite{blundell2015weight} and a bottleneck layer of 15 units, which makes the dimension $R^{(0)}$ four times smaller than the feature dimension of 60, and finally the linear regression layer.
The weight and factor loading matrices were updated using Adam \cite{kingma2014adam} and the cost function was the MSE.
The likelihood \eqref{eq:lin} in the experiments was calculated using bias and diagonal covariance matrix.

A set of experiments was performed using Theano \cite{Bastien-Theano-2012} to evaluate the model using gender dependent and phrase independent UBMs.
We compared the DNN which updates the last layer using \eqref{eq:11} to the TF2-DNN which also updates the last layer and in addition includes speaker and session factors.
The results in Table \ref{tab:1} show both DNN systems performing under $1\%$ EER for the male and female tasks, but the performance is greater in the case of models using tied factors in the model.
In the paper we exposed some parallelism between the DNN and a GMM both adapted with MAP. 
We can see in the experiments that the range of EER achieved is also comparable to GMMs in other works \cite{zeinali2016deep}.
And the relative improvement provided by the TF2-DNN with respect to the DNN system is also similar to  \cite{Miguel2014}, although in that case the files were processed at 8kHz.

%

In RSR2015, phrase dependent UBMs can be more specific, but there are less data available for the UBM, which makes difficult to train a DNN with many layers.
For that scenario we propose the use of dropout \cite{srivastava14a} and to approximate the likelihoods using \eqref{eq:sam}.
We performed some experiments using the female subset and phrase dependent UBMs.
A dropout layer was interleaved in the encoder and the decoder, with $p = 0.05$. The TF2-DNN had as dimensions  $R^{(1)}=5$ and $R^{(2)}=20)$ and the system provided a $0.11\%$ of EER.
This preliminary experiment showed us that dropout and other Bayesian techniques can mitigate part of the effect of overfitting of large DNNs when learning small datasets, and the system still can provide well calibrated scores in the context of these models.


\begin{table}[tb]
\begin{center}
\caption{ \footnotesize Experimental results on RSR2015 part I  \cite{Larcher2014} impostor-correct, showing EER$\%$ and NIST 2008 and 2010 min costs.}
\label{tab:1}
\footnotesize         
\begin{tabularx}{0.47\textwidth}{>{\hsize=0.35\hsize}X>{\hsize=0.12\hsize}X>{\hsize=0.12\hsize}X>{\hsize=0.16\hsize}X>{\hsize=0.17\hsize}X>{\hsize=0.17\hsize}X>{\hsize=0.17\hsize}X>{\hsize=0.17\hsize}X>{\hsize=0.17\hsize}X}
\toprule
	        &       &  &  	   & Male  & &     \\
	 \cmidrule(r){4-6}
	 \cmidrule(r){7-9}
System & $R^{(1)}$ & $R^{(2)}$ & EER$\%$ & det08 & det10 \\
	 \midrule 
	 DNN 
	 		& - & - & 0.65  & 0.037 & 0.155   \\
	 \midrule 
	 TF2-DNN 	
	 		&   15  & 50  & 0.25 & 0.016 & 0.086  \\
            &       & 75  & 0.31 & 0.017 & 0.080  \\
            &       & 100 & 0.30 & 0.017 & 0.085  \\
            &   25  & 50  & 0.29 & 0.016 & 0.075  \\
	     	&       & 75  & \textbf{0.25}  & 0.015 & 0.075   \\
	 	    &       & 100 & 0.31  & 0.017 & 0.069   \\

	        &       &  &  	   & Female  &    \\
	 \cmidrule(r){4-6}
	 \cmidrule(r){7-9}
     & $R^{(1)}$ & $R^{(2)}$ & EER$\%$ & det08 & det10\\
	 \midrule 
	 DNN 
	 		& - & - & 0.50 & 0.021  & 0.084  \\
	 \midrule 
	 TF2-DNN 	
	 		&   15  & 50  & 0.17  &  0.006 & 0.019  \\
            &       & 75  & 0.16  &  0.006 & 0.026  \\
            &       & 100 &  0.17  &  0.006 & 0.030  \\
            &   25  & 50  &  0.15 &  0.007 & 0.028  \\
	     	&       & 75  & \textbf{0.13} &  0.006 & 0.028  \\
	 	    &       & 100 & 0.19 &  0.007 & 0.021  \\            
\bottomrule 
\end{tabularx}
\end{center}
\vspace{-8mm}
\end{table}

\vspace{-0.5mm}
\section{Conclusions}
\label{s5}

In this paper we present an end-to-end method for speaker recognition based on neural networks, using tied hidden variables to model speaker and session variability and a MAP and Bayesian techniques to enrol and evaluate trials.
The last layer of the network is defined as a linear regression probabilistic model that  can be adapted during the enrolment so that the model can calculate likelihood ratios to decide the trial evaluations.
To estimate the model parameters and the hidden variables a two-step backpropagation algorithm is used.
We have tested the models in the text dependent speaker recognition database RSR2015 part I providing competitive results with respect to previous approaches.
\vspace{-0.5mm}
\section{Acknowledgements}
This work is supported by the Spanish Government and European Union (project TIN2014--54288--C4--2--R), and by the European Commission 
 FP7 IAPP Marie Curie Action GA--610986. We gratefully acknowledge the support of NVIDIA Corporation with the donation of a Titan X GPU used for this research.

\bibliographystyle{IEEEtran_nourl}

\bibliography{paper}

\begin{thebibliography}{10}
\providecommand{\url}[1]{#1}
\csname url@samestyle\endcsname
\providecommand{\newblock}{\relax}
\providecommand{\bibinfo}[2]{#2}
\providecommand{\BIBentrySTDinterwordspacing}{\spaceskip=0pt\relax}
\providecommand{\BIBentryALTinterwordstretchfactor}{4}
\providecommand{\BIBentryALTinterwordspacing}{\spaceskip=\fontdimen2\font plus
\BIBentryALTinterwordstretchfactor\fontdimen3\font minus
  \fontdimen4\font\relax}
\providecommand{\BIBforeignlanguage}[2]{{%
\expandafter\ifx\csname l@#1\endcsname\relax
\typeout{** WARNING: IEEEtran.bst: No hyphenation pattern has been}%
\typeout{** loaded for the language `#1'. Using the pattern for}%
\typeout{** the default language instead.}%
\else
\language=\csname l@#1\endcsname
\fi
#2}}
\providecommand{\BIBdecl}{\relax}
\BIBdecl

\bibitem{Yin2007}
\BIBentryALTinterwordspacing
S.-C. Yin, R.~Rose, and P.~Kenny, ``{A Joint Factor Analysis Approach to
  Progressive Model Adaptation in Text-Independent Speaker Verification},''
  \emph{IEEE Transactions on Audio, Speech and Language Processing}, vol.~15,
  no.~7, pp. 1999--2010, Sep. 2007.
\BIBentrySTDinterwordspacing

\bibitem{Kenny2008}
\BIBentryALTinterwordspacing
P.~Kenny, P.~Ouellet, N.~Dehak, V.~Gupta, and P.~Dumouchel, ``{A Study of
  Interspeaker Variability in Speaker Verification},'' \emph{IEEE Transactions
  on Audio, Speech, and Language Processing}, vol.~16, no.~5, pp. 980--988,
  Jul. 2008.
\BIBentrySTDinterwordspacing

\bibitem{kenny2014joint}
P.~Kenny, T.~Stafylakis, J.~Alam, P.~Ouellet, and M.~Kockmann, ``Joint factor
  analysis for text-dependent speaker verification,'' in \emph{Proc. Odyssey
  Workshop}, 2014, pp. 1--8.

\bibitem{Miguel2014}
A.~Miguel, A.~Ortega, E.~Lleida, and C.~Vaquero, ``Factor analysis with
  sampling methods for text dependent speaker recognition.''\hskip 1em plus
  0.5em minus 0.4em\relax Singapore: Proc. Interspeech, 2014, pp.
  1342--–1346.

\bibitem{Stafylakis2016SpeakerRecognition}
\BIBentryALTinterwordspacing
T.~Stafylakis, P.~Kenny, M.~J. Alam, and M.~Kockmann, ``{Speaker and Channel
  Factors in Text-Dependent Speaker Recognition},'' \emph{IEEE/ACM Transactions
  on Audio, Speech, and Language Processing}, vol.~24, no.~1, pp. 65--78, 1
  2016.
\BIBentrySTDinterwordspacing

\bibitem{Dehak2010}
N.~Dehak, P.~Kenny, R.~Dehak, P.~Dumouchel, and P.~Ouellet, ``{Front-End Factor
  Analysis For Speaker Verification},'' \emph{Audio, Speech, and Language
  Processing, IEEE Transactions on}, vol.~19, no.~14, pp. 788--798, May 2010.

\bibitem{villalba-bay2cov-is2011}
J.~Villalba and N.~Br\"{u}mmer, ``{Towards Fully Bayesian Speaker Recognition:
  Integrating Out the Between-Speaker Covariance},'' in \emph{Interspeech
  2011}, Florence, 2011, pp. 28--31.

\bibitem{Stafylakis2013}
T.~Stafylakis, P.~Kenny, P.~Ouellet, J.~Perez, M.~Kockmann, and P.~Dumouchel,
  ``Text-dependent speaker recogntion using plda with uncertainty
  propagation,'' in \emph{Proc. Interspeech}, Lyon, France, August 2013.

\bibitem{Liu2015DeepVerification}
\BIBentryALTinterwordspacing
Y.~Liu, Y.~Qian, N.~Chen, T.~Fu, Y.~Zhang, and K.~Yu, ``{Deep feature for
  text-dependent speaker verification},'' \emph{Speech Communication}, vol.~73,
  no.~C, pp. 1--13, 10 2015.
\BIBentrySTDinterwordspacing

\bibitem{zeinali2016deep}
H.~Zeinali, L.~Burget, H.~Sameti, O.~Glembek, and O.~Plchot, ``Deep neural
  networks and hidden markov models in i-vector-based text-dependent speaker
  verification,'' in \emph{Odyssey-The Speaker and Language Recognition
  Workshop}, 2016.

\bibitem{Dey2016DeepVerification}
\BIBentryALTinterwordspacing
S.~Dey, S.~Madikeri, M.~Ferras, and P.~Motlicek, ``{Deep neural network based
  posteriors for text-dependent speaker verification},'' in \emph{2016 IEEE
  International Conference on Acoustics, Speech and Signal Processing
  (ICASSP)}.\hskip 1em plus 0.5em minus 0.4em\relax IEEE, 3 2016, pp.
  5050--5054.
\BIBentrySTDinterwordspacing

\bibitem{zeinali2016vector}
H.~Zeinali, H.~Sameti, L.~Burget, J.~Cernocky, N.~Maghsoodi, and P.~Matejka,
  ``i-vector/hmm based text-dependent speaker verification system for reddots
  challenge,'' in \emph{Proc. Interspeech}.\hskip 1em plus 0.5em minus
  0.4em\relax ISCA, 2016.

\bibitem{Moreno2016}
G.~Heigold, I.~Moreno, S.~Bengio, and N.~M. Shazeer, ``End-to-end
  text-dependent speaker verification,'' in \emph{Proc. ICASSP}, 2016.

\bibitem{garimella2013factor}
S.~Garimella and H.~Hermansky, ``Factor analysis of auto-associative neural
  networks with application in speaker verification,'' \emph{IEEE transactions
  on neural networks and learning systems}, vol.~24, no.~4, pp. 522--528, 2013.

\bibitem{kingma2013auto}
D.~P. Kingma and M.~Welling, ``Auto-encoding variational bayes,'' in
  \emph{Proc. ICLR}, no. 2014, 2013.

\bibitem{kingma2015variational}
D.~P. Kingma, T.~Salimans, and M.~Welling, ``Variational dropout and the local
  reparameterization trick,'' in \emph{Proc. NIPS}, 2015.

\bibitem{blundell2015weight}
C.~Blundell, J.~Cornebise, K.~Kavukcuoglu, and D.~Wierstra, ``Weight
  uncertainty in neural network,'' in \emph{Proc. ICML}, 2015, pp. 1613--1622.

\bibitem{Moreno2014}
A.~Senior and I.~Lopez-Moreno, ``Improving dnn speaker independence with
  i-vector inputs,'' in \emph{Proc. ICASSP}, 2014.

\bibitem{xue2014fast}
S.~Xue, O.~Abdel-Hamid, H.~Jiang, L.~Dai, and Q.~Liu, ``Fast adaptation of deep
  neural network based on discriminant codes for speech recognition,''
  \emph{IEEE/ACM Transactions on Audio, Speech and Language Processing},
  vol.~22, no.~12, pp. 1713--1725, 2014.

\bibitem{samarakoon2016factorized}
L.~Samarakoon and K.~C. Sim, ``Factorized hidden layer adaptation for deep
  neural network based acoustic modeling,'' \emph{IEEE/ACM Transactions on
  Audio, Speech, and Language Processing}, vol.~24, no.~12, pp. 2241--2250,
  2016.

\bibitem{huang2016speaker}
Z.~Huang, J.~Tang, S.~Xue, and L.~Dai, ``Speaker adaptation of rnn-blstm for
  speech recognition based on speaker code,'' in \emph{Proc. ICASSP}.\hskip 1em
  plus 0.5em minus 0.4em\relax IEEE, 2016, pp. 5305--5309.

\bibitem{xue2015unsupervised}
S.~Xue, H.~Jiang, L.~Dai, and Q.~Liu, ``Unsupervised speaker adaptation of deep
  neural network based on the combination of speaker codes and singular value
  decomposition for speech recognition,'' in \emph{Proc. ICASSP}.\hskip 1em
  plus 0.5em minus 0.4em\relax IEEE, 2015, pp. 4555--4559.

\bibitem{Prince2008TiedDifferences}
\BIBentryALTinterwordspacing
S.~Prince, J.~Warrell, J.~Elder, and F.~Felisberti, ``{Tied Factor Analysis for
  Face Recognition across Large Pose Differences},'' \emph{IEEE Transactions on
  Pattern Analysis and Machine Intelligence}, vol.~30, no.~6, pp. 970--984, 6
  2008.
\BIBentrySTDinterwordspacing

\bibitem{Kuhn2000}
\BIBentryALTinterwordspacing
R.~Kuhn, J.-C. Junqua, P.~Nguyen, and N.~Niedzielski, ``{Rapid speaker
  adaptation in eigenvoice space},'' \emph{IEEE Transactions on Speech and
  Audio Processing}, vol.~8, no.~6, pp. 695--707, 2000.
\BIBentrySTDinterwordspacing

\bibitem{Kenny2005a}
\BIBentryALTinterwordspacing
P.~Kenny, G.~Boulianne, and P.~Dumouchel,
  ``\BIBforeignlanguage{English}{{Eigenvoice modeling with sparse training
  data}},'' \emph{\BIBforeignlanguage{English}{IEEE Transactions on Speech and
  Audio Processing}}, vol.~13, no.~3, pp. 345--354, May 2005.
\BIBentrySTDinterwordspacing

\bibitem{Martinez2011a}
D.~Mart\'{\i}nez, O.~Plchot, L.~Burget, G.~Ondrej, and P.~Matejka, ``{Language
  Recognition in iVectors Space},'' in \emph{Proc. Interspeech}, Florence,
  Italy, 2011.

\bibitem{Martinez2013a}
D.~Mart\'{\i}nez, E.~Lleida, A.~Ortega, and A.~Miguel, ``{Prosodic Features and
  Formant Modeling for an iVector-Based Language Recognition System},'' in
  \emph{ICASSP}, Vancouver, Canada, 2013.

\bibitem{Vaquero2013}
\BIBentryALTinterwordspacing
C.~Vaquero, A.~Ortega, A.~Miguel, and E.~Lleida, ``{Quality Assessment for
  Speaker Diarization and Its Application in Speaker Characterization},''
  \emph{IEEE Transactions on Audio, Speech, and Language Processing}, vol.~21,
  no.~4, pp. 816--827, Apr. 2013.
\BIBentrySTDinterwordspacing

\bibitem{Castan2013b}
D.~Castan, A.~Ortega, J.~Villalba, A.~Miguel, and E.~Lleida,
  ``{SEGMENTATION-BY-CLASSIFICATION SYSTEM BASED ON FACTOR ANALYSIS},'' in
  \emph{IEEE International Conference on Acoustics, Speech, and Signal
  Processing (ICASSP)}, 2013.

\bibitem{Kenny2006}
P.~Kenny, ``{Joint Factor Analysis of Speaker and Session Variability : Theory
  and Algorithms},'' {CRIM, Montreal, CRIM-06/08-13}, Tech. Rep., 2005.

\bibitem{gelman2006data}
\BIBentryALTinterwordspacing
A.~Gelman and J.~Hill, \emph{Data Analysis Using Regression and
  Multilevel/Hierarchical Models}, ser. Analytical Methods for Social
  Research.\hskip 1em plus 0.5em minus 0.4em\relax Cambridge University Press,
  2006.
\BIBentrySTDinterwordspacing

\bibitem{Tran2017}
\BIBentryALTinterwordspacing
D.~Tran, R.~Ranganath, and D.~M. Blei, ``{Deep and Hierarchical Implicit
  Models},'' Feb. 2017. [Online]. Available:
  \url{http://arxiv.org/abs/1702.08896}
\BIBentrySTDinterwordspacing

\bibitem{GhahramaniVariationalAnalysersb}
Z.~Ghahramani and M.~J. Beal, ``{Variational Inference for Bayesian Mixtures of
  Factor Analysers}.''

\bibitem{Ghahramani96}
\BIBentryALTinterwordspacing
Z.~Ghahramani and G.~Hinton, ``{The EM algorithm for mixtures of factor
  analyzers},'' Dept. of Comp. Sci., Univ. of Toronto, Toronto, Tech. Rep.~1,
  1996. [Online]. Available:
  \url{http://www.learning.eng.cam.ac.uk/zoubin/papers/tr-96-1.pdf}
\BIBentrySTDinterwordspacing

\bibitem{Bishop2008ALearning}
C.~M. Bishop, ``{A New Framework for Machine Learning},'' pp. 1--24, 2008.

\bibitem{welling2011bayesian}
M.~Welling and Y.~W. Teh, ``Bayesian learning via stochastic gradient langevin
  dynamics,'' in \emph{Proc. ICML}, 2011, pp. 681--688.

\bibitem{srivastava14a}
\BIBentryALTinterwordspacing
N.~Srivastava, G.~Hinton, A.~Krizhevsky, I.~Sutskever, and R.~Salakhutdinov,
  ``Dropout: A simple way to prevent neural networks from overfitting,''
  \emph{Journal of Machine Learning Research}, vol.~15, pp. 1929--1958, 2014.
\BIBentrySTDinterwordspacing

\bibitem{Larcher2014}
A.~Larcher, K.~A. Lee, B.~Ma, and H.~Li, ``{Text-dependent Speaker
  Verification: Classifiers, databases and RSR2015},'' \emph{Speech
  Communication}, vol.~60, pp. 56--77, 2014.

\bibitem{kingma2014adam}
D.~P. Kingma and J.~Ba, ``Adam: A method for stochastic optimization,'' in
  \emph{Proc. ICLR}, 2014.

\bibitem{Bastien-Theano-2012}
F.~Bastien, P.~Lamblin, R.~Pascanu, J.~Bergstra, I.~J. Goodfellow, A.~Bergeron,
  N.~Bouchard, and Y.~Bengio, ``Theano: new features and speed improvements,''
  Deep Learning and Unsupervised Feature Learning NIPS 2012 Workshop, 2012.

\end{thebibliography}

\end{document}